\documentclass[11pt, a4paper]{article}

\usepackage{amsmath,amssymb,amsthm}
\usepackage{url,eurosym}
\usepackage{graphicx,color} 
\usepackage{subcaption}
\usepackage{tabularx}

\newcommand{\rr}{\mathbb{R}}

\newcommand{\eqdef}{\triangleq}

\newcommand{\one}{\mathbf{1}}

\newcommand{\beqar}{\begin{eqnarray}}
\newcommand{\eeqar}{\end{eqnarray}}
\newcommand{\beqarno}{\begin{eqnarray*}}
\newcommand{\eeqarno}{\end{eqnarray*}}
\newcommand{\ba}[1]{\begin{array}{#1}}
\newcommand{\ea}{\end{array}}

\newcommand{\col}{\mathop{\rm col}\nolimits}

\begin{document}

	\title{NashOpt - A Python Library for Computing Generalized Nash Equilibria}
	
	\author{Alberto Bemporad 
    \thanks{The author is with the IMT School for Advanced Studies, Piazza San Francesco 19, Lucca, Italy. Email: \texttt{\scriptsize alberto.bemporad@imtlucca.it}.
    This work was funded by the European Union (ERC Advanced Research Grant COMPACT, No. 101141351). Views and opinions expressed are however those of the authors only and do not necessarily reflect those of the European Union or the European Research Council. Neither the European Union nor the granting authority can be held responsible for them.
}}
	
\maketitle
\thispagestyle{empty}
\begin{abstract}
\texttt{NashOpt} is an open-source Python library for computing and designing generalized Nash equilibria (GNEs) in noncooperative games with shared constraints and real-valued decision variables. The library exploits the joint Karush-Kuhn-Tucker (KKT) conditions of all players to handle both general nonlinear GNEs and linear–quadratic games, including their variational versions. Nonlinear games are solved via nonlinear least-squares formulations, relying on JAX for automatic differentiation. Linear-quadratic GNEs are reformulated as mixed-integer linear programs, enabling efficient computation of multiple equilibria. The framework also supports inverse-game and Stackelberg game-design problems. The capabilities of \texttt{NashOpt} are demonstrated through several examples, including noncooperative game-theoretic control problems of linear quadratic regulation and model predictive control. The library is available at \url{https://github.com/bemporad/nashopt}.
\end{abstract}

\section{Introduction}
Generalized games describe situations where several agents take decisions by optimizing individual objective functions under local and coupling constraints on their decision variables. Examples include traffic participants using the same roads, wireless systems sharing limited bandwidth, energy systems where prosumers share transmission capacity and balance constraints~\cite{LJWA20,WLMCMWW21}, or firms competing in a market~\cite{BYGP22}. The solution concept for these problems, in which the cost function and the feasible set of each agent depends on the decisions taken by the other agents, is the generalized Nash equilibrium (GNE), 
a state in which no agent has an incentive or the possibility to change their decision~\cite{Bau16}.

GNEPs have been widely studied in the literature, with various algorithms proposed for their solution, including splitting-based operator frameworks~\cite{YP19,BYGP22}, approaches built on augmented Lagrangian formulations~\cite{PF05,KS16,BFH19,LSM22}, interior-point optimization strategies~\cite{DFKS11}, Newton-type algorithms~\cite{FFP07}, distributed algorithms~\cite{YP19,TSN21,BYGP22}, and learning-based methods only relying on measuring best responses~\cite{FB25}. For the specific class of GNEPs in which the shared constraints are linear, efficient solution methods exist for linear objectives~\cite{DS16,GKW25}, multiparametric quadratic programming for strictly convex quadratic objectives~\cite{HB26}, and distributed payoff-based algorithms for convex games~\cite{TK18}.

\subsection{Contribution}
Motivated by the relative scarcity of software packages for solving rather general classes of GNEPs and their application to game-theoretic control problems, in this paper we describe the mathematical formulations of our package \texttt{NashOpt}, a Python library for solving GNEPs based on satisfying the KKT conditions of each player jointly. The library supports nonlinear GNEPs, which are solved via nonlinear least-squares methods, and 
linear-quadratic GNEPs, which can be solved more efficiently via mixed-integer linear programming (MILP) and can be used to enumerate multiple equilibria for the same game corresponding to different combinations of active constraints at optimality. We also address 
game-design problems, where a central authority can optimize some parameters of the game to achieve a desired equilibrium   in accordance with the {optimistic} Single-Leader-Multi-Follower (SLMF)
Stackelberg game setting~\cite{AS20}. Moreover, we leverage the GNE solution methods to solve game-theoretic control problems, including non-cooperative linear quadratic regulation (LQR) and model predictive control (MPC) problems~\cite{HBLD22,LK24}. The main purpose of the tool is to analyze and predict the effect of different agents pursuing their own objectives in a shared environment, and to design the game to promote desired equilibrium behaviors.

For nonlinear GNEs, we leverage JAX~\cite{JAX} for automatic differentiation and just-in-time compilation to achieve high performance, and on the open-source MILP solver HiGHS~\cite{HH18} or Gurobi~\cite{Gurobi}
for the linear-quadratic case, using the so-called ``big-M'' formulation to encode the complementary slackness conditions of the KKT system~\cite{HDB01a,BFSS25}.

The library is open-source and can be easily installed via \texttt{pip install nashopt}. {We refer the reader to the online documentation at \url{https://github.com/bemporad/nashopt} for a detailed description of the syntax and usage of the library.}

\subsection{Notation}
We denote by $\rr$ the set of real numbers. Given $N$ vectors $x_i\in\rr^{n_i}$, $i=1,\ldots,N$, we denote by $x=\col(x_1,\ldots,x_N)\in\rr^n$ their vertical concatenation, where $n=\sum_{i=1}^N n_i$. We
denote by $\one$ the vector of all ones of appropriate dimension, by $I$ the identity matrix, and by $e_i$ the $i$th column of $I$. Given a matrix $A$, we denote by $A^\top$ its transpose and by $\|A\|_F$ its Frobenius norm.

\section{Problem Formulation}
Consider the following game-design problem based on a multiparametric 
generalized Nash equilibrium problem (GNEP) with $N$ players:
\begin{equation}
    \begin{array}{rrl}
    \min_{x^\star, p\in P} &J({x^\star},p)\\
    \text{s.t. } & {x_i^\star}\in\arg\min_{x_i} &f_i(x,p)\\
    &\text{s.t. } & g(x,p) \leq 0\\
    &&h(x,p)=0 \\
    &&\ell_i \leq x\leq u_i \\
    &&x_{-i} = {x_{-i}^\star}\\
    &&i = 1, \ldots, N\\
    \end{array}
\label{eq:mpGNEP}
\end{equation}
where $x_i\in\rr^{n_i}$ collects the decision variables optimized by agent $i$,
$i=1,\ldots,N$, $x_{-i}$ collects those of all other agents,
$x = \col(x_1,\ldots,x_N) \in \rr^n$ is the 
stacked vector of all agents' decision variables, 
{$x^\star$ is constrained to be a} generalized Nash equilibrium (GNE) of the game {corresponding to the} parameter vector $p\in\rr^{n_p}$, $n_p\geq 0$, where by $n_p=0$ we mean that
no parameter enters~\eqref{eq:mpGNEP}. In~\eqref{eq:mpGNEP}, $f_i:\rr^n\to\rr$ is the objective of agent $i$, $g:\mathbb{R}^n \to \mathbb{R}^{n_g}$ 
and $h:\mathbb{R}^n \to \mathbb{R}^{n_h}$ define, respectively, {shared} inequality and equality constraints, 
and $\ell_i\in(\rr\cup\{-\infty\})^{n_i}$, $u_i\in(\rr\cup\{+\infty\})^{n_i}$ define the (local) box constraints for each player $i$. Moreover, $J:\mathbb{R}^n\times\mathbb{R}^{n_p}\to\mathbb{R}$ is the objective function used to shape the desired equilibrium of the game,
and $P\subseteq\mathbb{R}^{n_p}$ is a given feasible set for the parameter vector $p$,
such as the hyper-box $P=\{p\in \mathbb{R}^{n_p}:\ \ell_p\leq p\leq u_p\}$ where $\ell_p,u_p$ are (possibly infinite) bounds.
Problem~\eqref{eq:mpGNEP} is also referred to as an optimistic equilibrium of Single-Leader-Multi-Follower (Stackelberg) game~\cite{AS20}, with the leader optimizing $p$ and the followers playing the GNEP {determined} by $p$.

{For any given $p$, a GNE $x^\star$} is a vector for which no agent can reduce their cost given the strategies of the other players and feasibility constraints, i.e.,
\[
 f_i(\col(x_i^\star,x_{-i}^\star),p)
\leq
f_i(\col(x_i,x_{-i}^\star),p)
\qquad 
\forall x_i \in X_i(x_{-i}^\star,p)
\]
where, with a slight abuse of notation, we denoted by $\col(x_i,x_{-i})$ the vector obtained by collecting the components of $x_i$ and $x_{-i}$ in the right order, and $X_i(x_{-i}^\star,p)\subseteq\mathbb{R}^{n_i}$ is the feasible set of player $i$ defined by the constraints in~\eqref{eq:mpGNEP} given the strategies $x_{-i}$ of the other players. We highlight special cases of~\eqref{eq:mpGNEP} in the next sections.

\subsection{Solving a specific GNEP}
To just solve a given game, we set $J(x^\star,p)\equiv 0$ and $P=\{p_0\}$ a singleton, leading to the standard GNEP:
\begin{equation}
    \begin{array}{rl}
    x_i^\star = \arg\min_{\ell_i\leq x_i\leq u_i} &f_i(x,p_0)\\
    \text{s.t. } & g(x,p_0) \leq 0,\ h(x,p_0)=0 \\
        & x_{-i}=x_{-i}^\star,\quad i = 1, \ldots, N.
    \end{array}
\label{eq:GNEP}
\end{equation}  

\subsection{Finding a parameter vector for which a GNE exists}
If no specific equilibrium is desired, we can set $J(x^\star,p)\equiv 0$ and let $P$ be a non-singleton set, leading to the problem of finding a parameter vector $p\in P$ for which an equilibrium exists.

\subsection{Inverse game problem}
Given an observed agents' equilibrium $x_{\textrm des}$ (or a desired one we wish to achieve), by setting $J(x^\star,p)=\|x^\star-x_{\rm des}\|_2^2$ (or $J(x^\star,p)=\|x^\star-x_{\rm des}\|_\infty$, 
$\|x^\star-x_{\rm des}\|_1$), we solve the {\it inverse game} of finding a vector $p$ (if one exists) such that $x^\star\approx x_{\textrm des}$.

\section{KKT Conditions and Nonlinear Least-Squares Formulation}
\label{sec:KKT}
Let us assume that $f_i$, $g$, and $h$ are continuously differentiable with respect to $x_i$ and that suitable constraint qualifications hold, such as the Linear Independence Constraint Qualification (LICQ) condition at the GNE $x^\star(p)$\footnote{The LICQ condition is verified if the gradients of the active constraints at $x^\star(p)$ 
\[
\bigl\{ \nabla_{x_i} g_j(x^\star,p) : j \in \mathcal{A}(x^\star) \bigr\}
\;\cup\;
\bigl\{ \nabla_{x_i} h_k(x^\star,p) \bigr\}
\;\cup\;
\bigl\{ \pm e_k : x_{i,k}^\star = \ell_{i,k} \text{ or } u_{i,k} \bigr\}
\]
are linearly independent, where $e_k$ is the $k$-th unit vector of appropriate dimension, $\mathcal{A}(x^\star) = \{ j : g_j(x^\star,p) = 0 \}$ is the set of active inequality constraints at $x^\star$, and $x_{i,k}$, $\ell_{i,k}$, and $u_{i,k}$ are the $k$-th components of $x_i$, $\ell_i$, and $u_i$, respectively.}. Then, the necessary KKT conditions of optimality for each player $i$ read as follows \cite[Theorem 12.1]{NW06}:
\begin{subequations}
\begin{align}
&\nabla_{x_i} f_i(x,p) + \nabla_{x_i} g(x,p)^\top \lambda_i + \nabla_{x_i} h(x,p)^\top\mu_i -v_i + y_i= 0
\label{eq:stationarity}
\\[6pt]
&h(x,p)=0\\
&g(x,p) \le 0,\ \ell \leq x \leq u
\label{eq:primal}
\\[6pt]
&\lambda_i \ge 0,\ v_i \ge 0,\ y_i \ge 0
\label{eq:dual}
\\[6pt]
&\lambda_{i,j}g_j(x,p) = 0,\ v_{i,k}(x_{i,k} - \ell_{i,k}) = 0,\ y_{i,k}(u_{i,k} - x_{i,k}) = 0
\label{eq:complementarity}\\
&j=1,\ldots,n_g,\ k=1,\ldots,n_i,\ i=1,\dots,N\nonumber
\end{align}
\label{eq:kkt}%
\end{subequations}%
where $\lambda_i \in \mathbb{R}^{m_i}$ denotes the vector of Lagrange multipliers 
of player $i$ associated with the shared inequality constraints $g(x,p) \le 0$, $\mu_i$ the multipliers corresponding to the shared equality constraints $h(x,p)=0$, and $v_i$, $y_i$ the multipliers associated with bound constraints on $x_i$. {Note that the problem setting in~\eqref{eq:mpGNEP} could be immediately extended to allow for {\it player-specific} constraints with coupling of the form $g_i(x,p)\leq 0$, $h_i(x,p)=0$ in the KKT system~\eqref{eq:kkt}.}  

A generalized Nash equilibrium is a point $x^\star$ for which the KKT system
\eqref{eq:kkt} holds \emph{for all players simultaneously}. The KKT conditions can be expressed as the following zero-finding problem:
\begin{equation}
\begin{array}{rcl}
0&=&\nabla_{x_i} f_i(x,p) + \nabla_{x_i} g(x,p)^\top \lambda_i +\nabla_{x_i} h(x,p)^\top\mu_i\\
0&=&h(x{,p})\\
0&=&\sqrt{\lambda_{i,j}^2 + g_j(x,p)^2} - \lambda_{i,j} + g_j(x,p)\\
0&=&\sqrt{v_{i,k}^2 + (\ell_{i,k}-x_{i,k})^2} - v_{i,k} + \ell_{i,k} - x_{i,k}\\
0&=&\sqrt{y_{i,k}^2 + (x_{i,k}-u_{i,k})^2} - y_{i,k} + x_{i,k} - u_{i,k}\\
&&k=1,\ldots,n_i,\ i=1,\dots,N,\ j=1,\ldots,n_g
\end{array}
\label{eq:kkt-residual}
\end{equation}
where the last three residuals encode the primal feasibility~\eqref{eq:primal}, dual feasibility~\eqref{eq:dual}, and complementary slackness conditions~\eqref{eq:complementarity}
using the Fischer-Burmeister nonlinear complementarity problem (NCP) function $\phi:\rr\to\rr$~\cite{Fischer1992,FFK98} componentwise
\[
    \phi(\alpha,\beta) \eqdef \sqrt{\alpha^2 + \beta^2} - \alpha - \beta = 0 \quad\Longleftrightarrow\quad \alpha\beta = 0,\ \alpha \geq 0,\ \beta \geq 0.
\]
By letting $z$ be the vector collecting $x$, $\{\lambda_i\}$, $\{\mu_i\}$, $\{v_i\}$, and $\{y_i\}$, we can express the joint KKT conditions as
\begin{equation}
    R(z,p) = 0
\label{eq:zero-finding}
\end{equation}
where $R(z,p)$ is the residual function defined in~\eqref{eq:kkt-residual}. In the sequel, we will denote by $\nu=\col(\lambda,\mu,v,y)$ the stacked vector of all dual variables, so that $z=\col(x,\nu)$ is the overall vector of primal and dual variables. Given a solution $z^\star(p)$ satisfying~\eqref{eq:zero-finding}, we will implicitly denote by $x^\star(p)$
the corresponding GNE is obtained by extracting from $z^\star(p)$ the components corresponding to the primal variables.  We will also denote by $R(z)$ the residual function when $p$ is omitted or fixed.

For conditions of existence of a GNE for a given $p$, such as nonemptiness and compactness of the set of feasible decision vectors, convexity assumptions on $f$, $g$, and linearity of $h$, and constraint qualifications, we refer the reader to~\cite{FZ10}. 
{We remark that only relying on the joint KKT conditions~\eqref{eq:kkt} to characterize GNEs would require additional regularity assumptions. In \texttt{NashOpt}, these conditions are not verified numerically by the software. Consequently, when the assumptions above are violated, solving the KKT system should be interpreted as a heuristic computational approach for computing stationary points of the joint KKT system that, however, are not guaranteed {\it a priori} to be GNEs. On the other hand, in the case a zero residual $R$ is achieved within a given tolerance at a certain $x^\star$, whether this corresponds to a GNE or not can be immediately checked {\it a posteriori} by verifying that, for each player $i$, the best responses $\bar x_i(x^\star_{-i},p)\approx x_i^\star$, or that
the associated value function $f_i(\col(\bar x_i,x^\star_{-i}),p)\approx f_i(x^\star,p)$.}

\subsection{Determining a GNE via nonlinear least-squares}
For a given parameter $p_0$, we can compute the corresponding $z^\star(p_0)$
by minimizing the squared residual
\begin{equation}
    z^\star(p_0)\in\arg\min_z \frac{1}{2}\| R(z,p_0) \|_2^2
\label{eq:NLS}
\end{equation}
that serves as a {\it merit function} for the zero-finding problem~\eqref{eq:zero-finding}~\cite{FFK98,DFKS11}. The nonlinear least-squares problem~\eqref{eq:NLS} can be solved via standard methods, such as Levenberg-Marquardt (LM) algorithms~\cite{Lev44,Mar63}, possibly exploiting sparsity in the Jacobian 
of $R(z,p_0)$ with respect to $z$.
If the minimum residual $R(z^\star(p_0),p_0)$ of the nonlinear least-squares problem~\eqref{eq:NLS} is zero (or, more realistically, below a given positive tolerance), then $z^\star(p_0)$ contains a GNE solution $x^\star(p_0)$ and the associated Lagrange multipliers for each agent's problem associated with the parameter vector $p_0$.

\subsection{Variational GNE}
Variational GNEs (vGNEs) are obtained by simply enforcing that the Lagrange multipliers associated with the shared constraints are the same for all players, i.e., by replacing $\{\lambda_i\}$ with a single vector $\lambda$ and $\{\mu_i\}$ with a single vector $\mu$, which further reduces the number of variables in the zero-finding problem~\eqref{eq:zero-finding}. 

\subsection{Game design}
To solve the game-design problem~\eqref{eq:mpGNEP}, we can embed the 
KKT conditions~\eqref{eq:kkt} in the optimization of $p$, leading to the following mathematical program with equilibrium constraints (MPEC)~\cite{LPR96,GLY15}:
\begin{equation}
\begin{array}{rl}
    \min_{p, z} \quad & J([I\ 0]z,p)\\
    \text{s.t. } & R(z,p) = 0{,\ p\in P}.
\end{array}
\label{eq:game-design-MPEC}
\end{equation}
Problem~\eqref{eq:game-design-MPEC} can be solved via {constrained} nonlinear programming methods with respect to both $z$ and $p$.

To avoid the complexity of dealing with the equilibrium constraints, in our approach we solve the game-design problem~\eqref{eq:mpGNEP} by relaxing them and solving instead the following nonlinear programming problem:
\begin{equation}
\begin{array}{rl}
    \min_{p, z} \quad & J([I\ 0]z,p) +  \frac{\rho}{2}\| R(z,p) \|_2^2\\
    \text{s.t. } & p\in P
\end{array}
\label{eq:game-design-relaxed}
\end{equation}
where $\rho>0$ is a penalty parameter. By letting $\rho\to\infty$, the solution of~\eqref{eq:game-design-relaxed} approaches that of~\eqref{eq:game-design-MPEC}. When $P$ is a hyper-box (with possibly infinite lower and/or upper bounds on some parameters), 
Problem~\eqref{eq:game-design-relaxed} can be solved very efficiently via bound-constrained nonlinear optimization methods, such as L-BFGS-B~\cite{BLNZ95}.

The choice of the initial point $(z,p)$ and of the penalty $\rho$ are crucial to achieve convergence to a meaningful solution of~\eqref{eq:game-design-relaxed}; in particular, excessively small values of $\rho$ may lead to solutions that are far from satisfying the KKT conditions, while very large values of $\rho$ may cause the numerical solver to ignore the game-design objective $J(x,p)$.
In practice, we can solve~\eqref{eq:game-design-relaxed} with a moderate value of $\rho$ 
to get a value $(\bar z^\star,\bar p^\star)$ and then refine the GNE by solving the nonlinear least-squares problem~\eqref{eq:NLS} with hot-start at $(\bar z^\star, \bar p^\star)$.

\subsection{Non-smooth regularization}
When designing a parameter vector $p$, or finding an equilibrium among many existing ones for a given $p_0$, we can add the regularization term 
\[
    J_{\rm reg}(x,p) = \alpha_1\|x\|_1 + \alpha_2\|x\|_2^2,\quad \alpha_1,\alpha_2\geq 0
\]
in the game-design objective $J$; for example, {we can promote sparsity of the GNE $x^\star$} by choosing a sufficiently large $\alpha_1$. 
When $\alpha_1>0$, due to the non-differentiability of the $\ell_1$-norm at zero, we can reformulate the problem by splitting $x$ into positive and negative parts, $x=x_p-x_m$, with $x_p,x_m\geq 0$ and apply the result in~\cite[Corollary 1]{Bem25} to minimize 
\begin{equation}
\begin{array}{rl}
    \min_{p,x_p,x_m,\nu} \quad & J(x_p-x_m,p) + \frac{\rho}{2}\| R(\col(x_p-x_m,\nu),p) \|_2^2\\
        &+\alpha_1\one^\top(x_p+x_m) + \alpha_2(\|x_p\|_2^2 + \|x_m\|_2^2)\\
    \text{s.t. } & p\in P,\quad x_p \geq 0,\ x_m \geq 0
\end{array}
\label{eq:game-design-relaxed-L1}
\end{equation}
where $\nu$ collects all dual variables.

In the special case we are just looking for a sparse GNE with no game-design objective (i.e., $J(x,p)\equiv 0$) and $\alpha_2>0$, by letting $\alpha_3\eqdef\sqrt{2\alpha_2}$, we have
\[
    \begin{aligned}
    J_{\rm reg}(x,p) = &{\alpha_1\one^\top(x_p+x_m) + \alpha_2(\|x_p\|_2^2+ \|x_m\|_2^2)}\\
    = &\frac{1}{2}\left\|\alpha_3 x_p + \frac{\alpha_1}{\alpha_3}\one \right\|_2^2 + 
        \frac{1}{2}\left\|\alpha_3 x_m + \frac{\alpha_1}{\alpha_3}\one\right\|_2^2  + \textrm{const.}
    \end{aligned}
\]
Then, we can equivalently solve
\begin{equation}
    \begin{aligned}
    \min_{p, x_p, x_m, \nu} \quad & \frac{1}{2}\left\|\begin{bmatrix}
                \alpha_3 x_p + \frac{\alpha_1}{\alpha_3}\one \\
                \alpha_3 x_m + \frac{\alpha_1}{\alpha_3}\one \\
                \sqrt{\rho}\ R(\col(x_p-x_m,\nu),p)
                \end{bmatrix}\right\|_2^2\\
    \text{s.t. } & p\in P,\ x_p \geq 0,\ x_m \geq 0.
    \end{aligned}
\label{eq:GNE-relaxed-L1}
\end{equation}
When $P$ is a hyper-box (with possibly infinite lower and/or upper bounds on some parameters),
or a singleton $\{p_0\}$ as a special case, Problem~\eqref{eq:game-design-relaxed-L1} can be solved as a bound-constrained nonlinear least-squares problem, such as via a trust-region reflective algorithm~\cite{CL96}.
Lower and upper bounds $\ell\leq x\leq u$ can be explicitly included in~\eqref{eq:game-design-relaxed-L1} and~\eqref{eq:GNE-relaxed-L1} by tightening the nonnegativity constraints on $x_p$ and $x_m$, i.e., by setting
\[
    \max\{0,\ell\}\leq x_p\leq \max\{0,u\},\qquad \max\{0,-u\}\leq x_m\leq \max\{0,-\ell\}.
\]
The approach can be immediately extended to the case $J(x,p)=\frac{1}{2}\|f_J(x,p)\|_2^2$ for some vector function $f_J:\rr^n\times\rr^{n_p}\to n_f$ by extending the residual vector in~\eqref{eq:GNE-relaxed-L1} with $f_J(x_p-x_m,p)$.

\section{Linear Quadratic Games}
Consider the following special case of~\eqref{eq:mpGNEP} where the agents' cost functions are quadratic and the constraints are linear:
\begin{equation}
\begin{array}{rrl}
    \min_{p, x^\star} \quad & J(x^\star,p)\\
    \text{s.t. } & x^\star_i\in \arg\min_{x_i} &f_i(x,p)=\frac{1}{2} x^\top Q^i x + (c^i + F^i p)^\top x \\
    & \text{s.t. } & A x \leq b + S p\\ 
    &&A_{\mathrm{eq}} x = b_{\mathrm{eq}} + S_{\mathrm{eq}} p \\
    &&\ell_i \leq x \leq u_i\\
    &&x_{-i} = x^\star_{-i}\\
    && i=1,\dots,N
\end{array}
\label{eq:LQ-game}
\end{equation}
where $x^\star$ is the generalized Nash equilibrium of the Linear-Quadratic (LQ) game with $N$ players, 
$(Q^i, c^i, F^i)$ define the quadratic cost function for agent $i$, with $Q^i=(Q^i)^\top\in\rr^{n\times n}$, 
and the diagonal block $Q^i_{ii} \succeq 0$, 
$c^i\in\rr^n$, $F^i\in\rr^{n\times n_p}$, matrices $A\in\rr^{n_g\times n}$, $b\in\rr^{n_g}$, $S\in\rr^{n_g\times n_p}$ define the shared linear inequality constraints,
$A_{\mathrm{eq}}\in\rr^{n_h\times n}$, $b_{\mathrm{eq}}\in\rr^{n_h}$, $S_{\mathrm{eq}}\in\rr^{n_h\times n_p}$ define the shared linear equality constraints.

We consider as game-design function $J(x,p)$, if any is given, the sum of convex piecewise affine functions
\begin{subequations}
\begin{equation}
    J_{\rm PWA}(x,p) = \sum_{j=1}^{n_J}\max_{k=1,\dots,n_j} D_{jk} x + E_{jk} p + h_{jk}
\label{eq:PWA_game_cost}
\end{equation}
or {the} convex quadratic function
\begin{equation}
    J_{\rm Q}(x,p) = \frac{1}{2} [x^T\ p^T] Q_J \begin{bmatrix}x\\p\end{bmatrix} + c_J^T \begin{bmatrix}x\\p\end{bmatrix}
\label{eq:quad-game-obj}
\end{equation}
where $Q_J=Q_J^\top\succeq 0$, $Q_J\neq 0$, $Q_J\in\rr^{(n+n_p)\times(n+n_p)}$, $c_J\in\rr^{n+n_p}$,
or the sum of both, i.e., 
\begin{equation}
    J(x,p) = J_{\rm PWA}(x,p) + J_{\rm Q}(x,p).
\label{eq:total_game_cost}
\end{equation}
\label{eq:game-design-cost}%
\end{subequations}
Special cases are: ($i$) $J_{\rm PWA}(x,p)=J_{\rm Q}(x,p)\equiv 0$ for no game design; ($ii$)
$J_{\rm PWA}(x,p)=\|x-x_{\rm des}\|_\infty$ (i.e., $n_J=1$, $n_1=2n$, $D_{1} = [I\ -I]^\top$, $E_{1}=0$, $h_{1} = \col(-x_{\rm des},x_{\rm des})$), or $J_{\rm PWA}(x,p)=\|x-x_{\rm des}\|_1$ (i.e., $n_J=n$, $n_j=2$, $D_{j,1} = e_j^\top$, $D_{j,2} = -e_j^\top$, $E_{j,1}=E_{j,2}=0$, $h_{j,1} = - x_{\rm des,j}$, $h_{j,2} = x_{\rm des,j}$), or $J_{\rm Q}(x,p)=\frac{1}{2}\|x-x_{\rm des}\|_2^2$ for inverse game problems; {($iii$)}
the {\it social welfare} quadratic objective $J_{\rm Q}(x,p)=\sum_{i=1}^N f_i(x,p)$.

For convenience, let us embed finite lower and upper bounds on $x$ into the inequality constraints by defining
\[
\bar A = \begin{bmatrix} A \\ I_{u} \\ -I_{\ell} \end{bmatrix}, \quad \bar b = \begin{bmatrix} b \\ u_{u} \\ -\ell_{\ell} \end{bmatrix}, \quad \bar S = \begin{bmatrix} S \\ 0 \\ 0 \end{bmatrix}
\]
where $I_{\ell}$ and $I_{u}$ collect the rows of the identity matrix corresponding, respectively, to finite upper and lower bound, and \(\ell_{\ell}\) and \(u_{u}\) are the corresponding bounds. Let $m$ be the total number of resulting inequality constraints.

The KKT conditions for each player's optimization problem are necessary and sufficient for optimality, due to the convexity of the cost functions and constraints, and can be used to characterize the generalized Nash equilibria of the game:
\begin{equation}
    \begin{aligned}
    &Q_{i}^i x + c^i_{i} + F^i_i p + \bar A^\top_i \lambda_i + A_{\mathrm{eq},i}^\top \mu_i = 0\\
    &\bar A x \le \bar b + \bar S p\\
    &A_{\mathrm{eq}} x = b_{\mathrm{eq}} + S_{\mathrm{eq}} p\\
    &\lambda_i \ge 0\\
    &\lambda_{i,j} (\bar A_j x - \bar b_j - \bar S_j p) = 0,\quad i=1,\dots,N
    \end{aligned}
\label{eq:Nash-KKT}
\end{equation}
where $Q^i_i$ collects the rows of $Q^i$ corresponding to player $i$'s variables, $\bar A_i$ collects the columns and rows of $\bar A$ where $x_i$ is involved, 
$A_{\mathrm{eq},i}$ is defined similarly, and $\lambda_i$, $\mu_i$, are the Lagrange multipliers associated with the inequality and equality constraints, respectively, where the variables in $x_i$ appear.
{Note that the LQ problem~\eqref{eq:LQ-game} could be extended to handle player-specific linear constraints with coupling of the form $A^ix\leq b^i+S^ip$, $A_{\mathrm{eq}}^i x = b_{\mathrm{eq}}^i + S_{\mathrm{eq}}^i p$, by changing~\eqref{eq:Nash-KKT} accordingly.}  

By stacking the KKT conditions \eqref{eq:Nash-KKT} for all players, we can rewrite the generalized Nash equilibrium conditions as a single mixed-integer linear program (MILP) using standard big-M constraints to model the complementarity conditions (see, e.g.,~\cite[Proposition 2]{HDB01a}):
\begin{equation}
\begin{array}{rl}
    \min_{p, x, \lambda, \mu, \delta, \sigma} \quad & \displaystyle{
    \frac{1}{2} [x^T\ p^T] Q_J \begin{bmatrix}x\\p\end{bmatrix} + c_J^T \begin{bmatrix}x\\p\end{bmatrix}} + \sum_{j=1}^{n_J}\sigma_j\\
    \text{s.t. } &\sigma_j\geq D_{jk} x + E_{jk} p + h_{jk}, \quad k=1,\dots,n_j \\
    & Q^i_i x + c^i_i + F^i_i p + \bar A^\top_i \lambda_i + A_{\mathrm{eq},i}^\top \mu_i = 0 \\
    &\bar A x \le \bar b + \bar S p\\ 
    &A_{\mathrm{eq}} x = b_{\mathrm{eq}} + S_{\mathrm{eq}} p \\
    &\lambda_i \ge 0,\ i=1,\ldots,N\\
    &\bar A x - \bar b - \bar S p \le M (1 - \delta) \\
    &\lambda_{i} \le M \delta,\ i=1,\ldots,N  \\
    & \delta \in \{0,1\}^m
\end{array}
\label{eq:GNE-MIP}
\end{equation}
where $M\in\rr$ is a sufficiently large positive constant, $\delta$ is a vector of binary variables introduced to model the complementarity conditions, and $\sigma$ is a vector of auxiliary variables used to define an upper-bound the objective function, $\sigma\in\rr^{n_J}$. Clearly, at optimality $\sum_{i=1}^{n_J}\sigma_j = J_{\rm PWA}(x,p)$, as the first $n_j$ inequalities in~\eqref{eq:GNE-MIP} become all active.
Note that using the same vector $\delta$ for all players ensures that shared constraints are active/inactive for all players simultaneously.

Problem~\eqref{eq:GNE-MIP} can be either solved 
with respect to $p, x, \lambda, \mu, \delta, \sigma$ by MILP when $J_{\rm Q}(x,p)\equiv 0$, or by mixed-integer quadratic programming (MIQP) otherwise. 

\subsection{Extracting multiple equilibria}
For the same parameter vector $p$, multiple generalized Nash equilibria may exist that correspond to different combinations $\bar\delta$ of active inequality constraints, as recently explicitly characterized in~\cite{HB26}. To extract multiple equilibria, we can iteratively solve the MILP above after adding the following ``no-good`` constraint 
\begin{equation}
    \sum_{i:\bar \delta_i=1} \delta_i - \sum_{i:\bar \delta_i=0} \delta_i \leq \left(\sum_{i=1}^m \bar \delta_i\right) - 1
\label{eq:no-good}
\end{equation}
which prevents the MILP from returning the same combination $\bar \delta$ of active constraints in all subsequent iterations.

\subsection{Variational GNE {of LQ games}}
As for the nonlinear case, a vGNE can be obtained by enforcing that the Lagrange multipliers associated with the shared constraints are the same for all players, which further reduces the number of variables in~\eqref{eq:GNE-MIP}.

{
\subsubsection{Proximal ADMM solver}
\label{sec:prox-admm}
For {vGNEs of} LQ games with fixed parameter $p=p_0$ and no game objective, we also consider the 
proximal-ADMM algorithm in~\cite{BK21} as an alternative approach to compute a vGNE. To apply the method, 
that only handles equality constraints, we reformulate the problem as the following LQ game with $N+1$ players:
\begin{equation}
    \begin{aligned}
    x_i^\star \in \arg\min_{x_i} &\frac{1}{2} x'Q^i x + (c^i)^T x\\
    \text{s.t. } &\bar A x + s^\star = \bar b, \quad A_{\mathrm{eq}} x = b_{\mathrm{eq}}\\
        & x_{-i} = x_{-i}^\star,\quad  i=1,\ldots,N\\
    s^\star \in \arg\min_{s} &\frac{1}{2}\left\|\bar A x^\star + s - \bar b\right\|_2^2\\
    \text{s.t. } &\bar A x^\star + s = \bar b, s\geq 0
    \end{aligned}
    \label{eq:prox-admm-reformulation}
\end{equation}
where, without loss of generality, we have taken $p_0=0$.
The proximal-ADMM iterations for the reformulation~\eqref{eq:prox-admm-reformulation} are as follows:
\begin{subequations}
\begin{align}
     x_i^{k+1} &\in \arg\min_{x_i} \frac{1}{2} x_i'Q^i_{ii} x_i + (c^i_i+Q^i_{i,-i} \bar x_{-i}^k+(\lambda^k)^\top \bar A_i + (\mu^k)^\top A_{\mathrm{eq},i})x_i + \frac{\gamma}{2}\|x_i - x_i^k\|_2^2\nonumber\\& + \frac{\rho}{2}(\|\bar A_i x_i^k + \bar A_{-i} \bar x_{-i}^k + s^k - \bar b\|_2^2+\|A_{\mathrm{eq},i} x_i + A_{\mathrm{eq},-i} \bar x_{-i}^k - d\|_2^2)\label{eq:prox-admm-x}\\
     s^{k+1} &\in \arg\min_{s\geq 0} \frac{1+\rho}{2}\left\|\bar A x^{k+1} + s - \bar b\right\|_2^2 + (\lambda^k)^\top s + \frac{\gamma}{2}\|s - s^k\|_2^2\label{eq:prox-admm-s}\\
     \lambda^{k+1} &= \lambda^k + \rho (\bar A x^{k+1} + s^{k+1} - \bar b)\\
     \mu^{k+1} &= \mu^k + \rho (A_{\mathrm{eq}} x^{k+1} - b_{\mathrm{eq}})
\end{align}    
\label{eq:prox-admm-algo}%
\end{subequations}
where $\bar x^k_{-i}$ collects the components of $x_1^{k+1},\ldots,x_{i-1}^{k+1},x_{i+1}^k,\ldots,x_N^k$, $\gamma>0$ is the proximal parameter, and $\rho$ is the augmented Lagrangian parameter. The iterations~\eqref{eq:prox-admm-algo} admit a closed-form solution: the $x_i$-update in~\eqref{eq:prox-admm-x} is an unconstrained strongly convex quadratic program (QP) that can be solved by caching the Cholesky decompositions of $Q^i_{ii}+\gamma I+\rho(\bar A_i^\top \bar A_i + A_{\mathrm{eq},i}^\top A_{\mathrm{eq},i})$ for each player $i$ and then, at each iteration $k$, solving two triangular linear systems; the evaluation of $s^{k+1}$ also admits the closed-form solution 
\[
    s^{k+1} = \frac{1}{1+\rho+\gamma}\max\left((1+\rho)(\bar b - \bar A x^{k+1}) 
              + \gamma s^k -\mu^k, 0\right)
\]
as, by completing the squares,~\eqref{eq:prox-admm-s} is equivalent to the projection of a vector onto the nonnegative orthant.
}

\section{Game-Theoretic Control}

\subsection{Game-theoretic linear quadratic regulation}
We consider a non-cooperative multi-agent linear-quadratic regulator (LQR) problem where $N$ agents control a shared discrete-time linear dynamical system 
\begin{equation}
\begin{aligned}
x(t+1) &= A x(t) + B u(t),\qquad u(t) = \col(u_1(t),\ldots,u_N(t))\\
y(t) &= C x(t)
\end{aligned}
\label{eq:linear-system}
\end{equation}
where $x(t) \in \mathbb{R}^{n_x}$ is the state vector at step $t=0,1,\ldots$, $y(t)$ is the output vector, and $u(t) \in \mathbb{R}^{n_u}$ is the input vector partitioned among the $N$ agents, with $u_i(t) \in \mathbb{R}^{n_i}$ being the input applied by agent $i$ at time $t$, and $n_u=\sum_{i=1}^Nn_i$. Each agent applies their control input $u_i(t)$ according to the state-feedback law
\[
u_i(t) = -K_i x(t)
\]
where $K_i \in \mathbb{R}^{n_i\times n_x}$ is the LQR gain corresponding to weight matrices $Q_i\in \mathbb{R}^{n_x\times n_x}\succeq 0$, such as $Q_i=C^\top Q_{yi} C$ with $Q_{yi}\succeq 0$,
and $R_i\in\mathbb{R}^{n_i}$, $R_i\succ 0$. We approximate the {desired LQR gain $K_i$ as a function of 
the other agents' gains $K_{-i}$ by running the following Riccati-based iterations 
\begin{equation}
    \begin{aligned}
        K_i^k &= (R_i + B_i^T X_i^k B_i)^{-1} B_i^T X_i^k (A - B_{-i} K_{-i})\\
        A_k &= A - B_{-i} K_{-i} - B_i K_i^k\\
        X_i^{k+1} &= Q_i + A_k^T X_i^k A_k + (K_i^k)^T R_i K_i^k
    \end{aligned}
\label{eq:approx-LQR-cost}
\end{equation}
from $X_i^0 = Q_i$ for $k=1,\ldots,N_{\rm LQR}$. Then, we} define each agent's cost function as the squared Frobenius norm of the deviation from its optimal gain $K_i^{N_{\rm LQR}}(K_{-i})$:
\[
f_i(K_i, K_{-i}) = {\frac{1}{2}\|K_i^{N_{\rm LQR}}}(K_{-i}) - K_i\|_F^2.
\]
Let $K = [K_1^\top \ \cdots \ K_N^\top]^\top$ denote the stacked feedback gain.
A Nash equilibrium is a feedback gain matrix $K^\star$ such that
\[
K_i^\star \in \arg\min_{K_i} f_i(K_i, K_{-i}^\star)
\qquad \forall i=1,\dots,N.
\]
The equilibrium can be computed as the solution of a Nash equilibrium
problem by using the nonlinear least-squares approach~\eqref{eq:NLS} (without parameter $p_0$),
using the centralized LQR solution $K_{\rm LQR}$ associated with $(A,B$, $\sum_i Q_i$, $\mathrm{blkdiag}(R_1,\dots,R_N))$ as initial guess. 
{Note that, in this case, the optimality conditions~\eqref{eq:kkt-residual} reduce to $\nabla_{K_i} f_i(K_i, K_{-i}) = K_i^{N_{\rm LQR}}(K_{-i}) - K_i =0$ for all $i=1,\ldots,N$, where with a slight abuse of notation, we considered $K_i$ as its vectorized version; hence, the nonlinear least-squares problem~\eqref{eq:NLS} reduces to minimizing the sum of the squared deviations from the optimal gains, i.e., $\sum_{i=1}^N \|K_i^{N_{\rm LQR}}(K_{-i}) - K_i\|_F^2$.}

{Alternatively, the coupled discrete-time algebraic Riccati equations can be solved using Riccati-based iterations as described in~\cite[Section III.B]{NMSM24} until convergence is reached within a given tolerance.}

\subsection{Game-theoretic model predictive control}
Consider again the discrete-time linear system~\eqref{eq:linear-system} with $N$ agents competing to make the output $y(t)$ track a given set-point $r(t)\in\mathbb{R}^{n_y}$. By letting $\Delta u(t)=u(t)-u(t-1)$ denote the input increment at time $t$, each agent $i$ chooses the sequence $\Delta u_i \triangleq \{\Delta u_{i,k}\}_{k=0}^{T-1}$ of input increments over a prediction horizon of $T$ by solving the following standard discrete-time finite-horizon linear optimal control problem~\cite{BMR04,Bem04b}:
\begin{equation}
\ba{rll}
(\Delta u_i,\epsilon_i) \in\arg\min & \multicolumn{2}{l}{\displaystyle{\left( \sum_{k=0}^{T-1}
(y_{k+1}-r(t))^\top {Q_{y,i}} (y_{k+1}-r(t))
      + \Delta u_{i,k}^\top Q_{\Delta u,i}\Delta u_{i,k}\right)}}\\
&\multicolumn{2}{l}{+ {q_{\epsilon,i}} \epsilon_i}\\
\text{s.t. } & x_{k+1} = A x_k + B u_k& \hspace*{-1cm}y_{k+1} = C x_{k+1}\\
& u_{k,i} = u_{k-1,i} + \Delta u_{k,i}& \hspace*{-1cm}u_{-1} = u(t-1)\\
&\Delta u_{\rm min} \leq \Delta u_k \leq \Delta u_{\rm max} 
& \hspace*{-1cm}u_{\rm min} \leq  u_k \leq u_{\rm max}\\
& y_{\min} - \sum_{i=1}^N \epsilon_i \leq y_{k+1} \leq y_{\max} + \sum_{i=1}^N \epsilon_i&
\epsilon_i \geq 0\\
& i=1,\ldots,N,\ k=0,\ldots,T-1. 
\ea
\label{eq:mpc-problem}
\end{equation}
In \eqref{eq:mpc-problem}, $Q_{\Delta u,i}\succ 0$ and ${Q_{y,i}}\succeq 0$; for example, if an agent is interested in controlling only a subset $I_i$ of the components of the output vector with unit weights, we have $Q_{y,i}=\sum_{j\in I_i}e_je_j^\top$. {The} slack variables $\epsilon_i\geq 0$ are
used to soften the shared output constraints (with linear penalty $q_{\epsilon,i}\geq 0$).

The predicted trajectories $\{x_k,y_k,u_k\}$ in~\eqref{eq:mpc-problem} satisfy the dynamic equations~\eqref{eq:linear-system} over the prediction horizon $k=0,\ldots,T$,
with initial condition $x_0=x(t)$ and $u_{-1}=u(t-1)$. Note that each agent optimizes the sequence of input increments $\Delta u_i$ and the slack variable $\epsilon_i$, so that constraints on inputs, input increments, and slacks are local to each agent, while output constraints are shared among all agents, each one softening them with possibly different slacks $\epsilon_i$.  Hence,
the resulting game-theoretic MPC problem is the GNE problem of finding sequences $\{\Delta u_i^\star\}_{i=1}^N$ (and slacks $\{\epsilon_i^\star\}_{i=1}^N$) optimizing~\eqref{eq:mpc-problem} for all agents $i=1,\ldots,N$ simultaneously.
In a receding-horizon fashion, only the first inputs $u_i(t)=u_i(t-1)+\Delta u_{i,0}^\star$ are applied for $i=1,\ldots,N$, and the entire procedure is repeated at $t+1$.

The number of constraints in each agent's problem~\eqref{eq:mpc-problem} can be reduced by imposing the constraints only on a shorter constraint horizon of $T_c<T$ steps. This can be especially useful when $T$ is large but we want to limit the number of Lagrange multipliers and binary variables involved in the MILP reformulation~\eqref{eq:GNE-MIP} of the GNEP, in which we remove $p$, $\sigma$, and the objective function~\eqref{eq:PWA_game_cost}, as we are only interested in computing the GNE of the noncooperative MPC problem.

If a \emph{variational} equilibrium is requested, the additional equalities enforcing
common Lagrange multipliers for the shared output constraints in \eqref{eq:mpc-problem} can be imposed.
In Section~\ref{sec:mpc-example}, we will consider an example of noncooperative MPC and show that it can deviate significantly from the {\it cooperative and centralized} MPC solution in which all moves are optimized jointly by solving a quadratic programming (QP) problem with decision variables $\Delta u = \col(\Delta u_1,\ldots,\Delta u_N)$, $\epsilon=\col(\epsilon_1,\ldots,\epsilon_N)$ and cost function given by the sum of all agents' costs in~\eqref{eq:mpc-problem}.

\section{Examples}
We report several numerical tests to illustrate the methods {described} in the previous sections for computing and designing generalized Nash equilibria. All tests are run on a Macbook Pro with Apple M4 Max chip and 64 GB RAM using JAX~\cite{JAX} for automatic differentiation and just-in-time compilation of the involved functions {when required}.

\subsection{Linear-quadratic game}
Let us first consider a simple linear-quadratic game with $N=3$ agents, each controlling two decision variables $x_i\in\mathbb{R}^2$, $i=1,2,3$, with cost functions
\begin{equation}
    f_i(x) = \frac{1}{2} x^\top x + i\ \one^\top x
\label{eq:LQ-cost}
\end{equation}
where $x=\col(x_1,x_2,x_3)\in\mathbb{R}^6$, and shared constraints
\[
\begin{bmatrix}
-0.4 & -0.1 & -2.1 &  1.6 & -1.8 & -0.8\\
 0.5 & -1.2 & -1.1 & -0.9 &  0.6 &  2.3\\
 0 & -1.1 &  0.5 & -0.6 &  0.0 &  1.2\\
-0.7 &  0 & -0.9 & -0.2 &  0.3 & -1
\end{bmatrix}x  \leq  \begin{bmatrix}
1\\
1\\
1\\
1
\end{bmatrix}.
\]
The GNE is computed by solving the MILP~\eqref{eq:GNE-MIP} with no parameter $p$,
and optimal combinations extracted by iteratively adding the no-good constraint~\eqref{eq:no-good},
leading to the three equilibrium solutions:
\[
\begin{aligned}
x^\star_1 &= [11.0588\  2.7647\ -1\     -1\     -2\    -2]^\top\\
x^\star_2 &= [ 0\      0\     -0.3436\ -1.5001\ -1.1599\ -0.7387]^\top\\
x^\star_3 &= [ 0\      0\     -0.7966\ -0.8336\ -0.2783\ -0.1998]^\top.
\end{aligned}
\]
These correspond to the optimal active constraint combinations $(1)$, $(1,4)$, and $(1,2,4)$, respectively. 
By imposing the additional constraint that the Lagrange multipliers associated with the shared constraints are the same for all agents, {we get} the vGNE
\[
{x^\star_v = [ 0.3553\  0.037\  0.0431\ -1.5324\ -1.4232\ -1.408 ]^\top}
\]
corresponding to the active constraint combination $(1,4)$. The CPU time for computing each equilibrium ranges between {$4.7$ and $10.1$}~ms for the non-variational case and is {$8.3$}~ms for the variational one using the HiGHS solver~\cite{HH18}. Note that infinitely many different GNEs may exist for the same combination of active constraints, as clearly shown in~\cite{HB26}. 

For comparison, we also use the nonlinear least-squares approach~\eqref{eq:NLS} and compute
the same vGNE $x^\star_v$ as above in {$229.9$}~ms (10 LM iterations) starting from the initial guess $x^{0}=0$. The same vGNE is found again by running the proximal ADMM approach described in Section~\ref{sec:prox-admm},
{where we set $\gamma=1$, $\rho=1$, $x^{0}=0$, $\lambda^{0}=0$, which takes {$13.9$}~ms (229 ADMM iterations). }

\begin{figure}[t]
    \centering
    \begin{subfigure}[t]{0.48\textwidth}
        \centering
        \includegraphics[width=\linewidth]{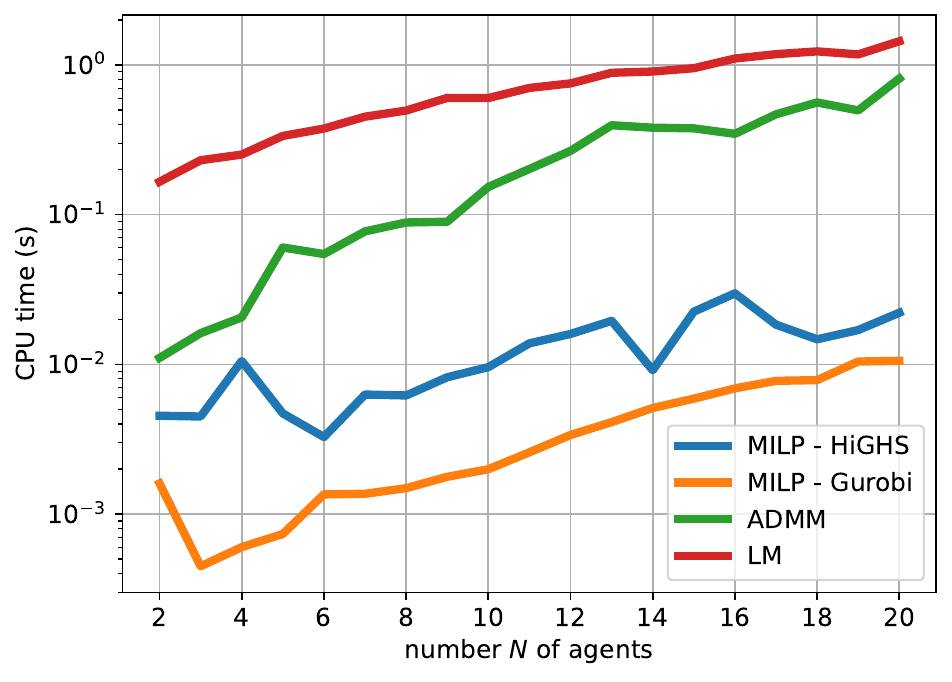}
        \caption{CPU time vs number $N$ of agents, under $n_g=2N$ shared constraints.}
        \label{fig:example_cputime_comparison}
    \end{subfigure}\hfill
    \begin{subfigure}[t]{0.48\textwidth}
        \centering
        \includegraphics[width=\linewidth]{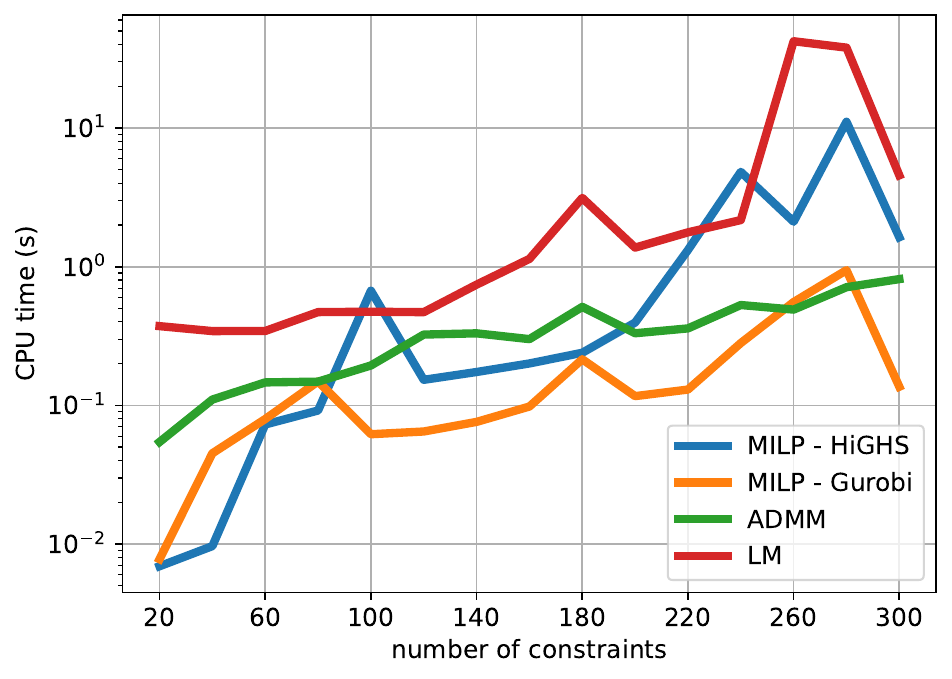}
        \caption{{CPU time vs number of shared constraints, with $N=5$ agents.}}
        \label{fig:example_cputime_comparison_constraints}
    \end{subfigure}
    \caption{CPU time for computing a GNE via MILP vs LM for increasing number of agents and constraints.}
    \label{fig:example_cputime_comparison_combined}
\end{figure}

Next, we further compare the MILP{, LM, and proximal ADMM} approaches on random GNE problems of increasing size.
We consider linear-quadratic games with $N$ agents, each controlling $n_i=2$ decision variables, with cost functions $f$ as in~\eqref{eq:LQ-cost} and shared constraints defined by $n_g=2N$ random linear inequalities with unit right-hand side. The results are shown in Figure~\ref{fig:example_cputime_comparison}, where{, for an increasing number $N$ of agents,} we report the CPU time for computing a GNE via MILP {and for computing a vGNE by LM and proximal ADMM}. Computing GNEs on LQ games via MILP {with HiGHS is} about two orders of magnitude more efficient than by LM, and even more by using Gurobi's state-of-the-art {MILP solver; it also outperforms proximal ADMM (which takes between 258 and 1794 iterations). Figure~\ref{fig:example_cputime_comparison_constraints} shows instead the CPU time taken when fixing $N=5$ agents and increasing the number of shared constraints $n_g$ up to 300, showing that MILP methods degrade more in performance compared to proximal ADMM due to the increasing number of binary variables in~\eqref{eq:GNE-MIP}}. 

\subsection{Inverse linear-quadratic game design}
We consider an inverse game-design problem for a GNEP with $N=10$ agents, each optimizing a vector $x_i\in\mathbb{R}^{n_i}$ of $n_i=10$ variables ($x\in\rr^{100})$, parameterized by a vector $p\in\mathbb{R}^{n_p}$, with $n_p=5$, with $\|p\|_\infty\leq 100$. For each given $p$, each agent $i=1,\ldots,N$ solves the convex quadratic optimization problem {defined in~\eqref{eq:LQ-game}} 
with $b\in\rr^{50}$, $b_{\mathrm{eq}}\in\rr^{5}$.
All matrices and vectors are randomly generated with appropriate dimensions, with $Q^i\succ 0$,
and the box constraints are defined by $u=-\ell=10\cdot\one$.

To formulate the inverse game problem in an ideal setting, we first compute a reference equilibrium $x_{\rm des}$ and a corresponding parameter vector $\hat p$ by solving~\eqref{eq:LQ-game} with zero objective 
$J(x^\star,p)=0$. Then, we solve the inverse game-design problem~\eqref{eq:GNE-MIP} by using either 
$J(x^\star,p)=\|x^\star - x_{\rm des}\|_\infty$ (MILP problem) or $J(x^\star,p)=\frac{1}{2}\|x^\star - x_{\rm des}\|_2^2$ (MIQP problem) to retrieve a parameter vector $p^\star$ that yields an equilibrium $x^\star(p^\star)\approx x_{\rm des}$. We use Gurobi's {solvers in both cases}.

The parameter vector $p^\star$ is retrieved, respectively, in $0.0394$~s by MILP 
with $\|x^\star(p^\star) - x_{\rm des}\|_\infty=5.4712\cdot 10^{-13}$,
and $0.0660$~s by MIQP with $\|x^\star(p^\star) - x_{\rm des}\|_2= 1.2434\cdot 10^{-14}$.

\subsection{Stackelberg game}
Consider a problem with $N$ agents (the followers), each controlling a decision variable $x_i\geq 0$, $i=1,\ldots,N$, subject to the shared constraint $\sum_{i=1}^N x_i \leq C$.
Each agent $i$ minimizes:
\[
f_i(x,p) = a_i x_i^2 + \sum_{j=1}^N \Gamma_{ij} x_i x_j + \pi_i(x,p) x_i
\]
where $a_i > 0$ is a self-cost coefficient, $\Gamma\in\rr^{N\times N}$ is a symmetric interaction matrix, and $\pi_i(x,p)$ is the nonlinear price function 
\[
    \pi_i(x,p)=p_{1,i}+p_2\left(\sum_{j=1}^N x_j\right)^2\!\!\!\!.
\]
Here $p=\col(p_{1,1},\ldots,p_{1,N},p_2)$ is the vector of game parameters chosen by the leader to minimize the loss function
\[
    J(x,p) = \left(\sum_{i=1}^N x - D\right)^2 + \eta \sum_{i=1}^N (p_{1,i} - \bar{p}_{1,i})^2-
\rho \sum_{i=1}^N \pi_i(x,p) x_i
\]
with $p_{1,i} \in [p_{1,\min},p_{1,\max}]$ and $p_2 \in [p_{2,\min},p_{2,\max}]$.
The numerical values are set as follows: $N=8$, $C=1$, $D=0.9$, $p_{1,\min}=-5$, $p_{1,\max}=0$,
$p_{2,\min}=0$, $p_{2,\max}=0.2$, $\bar{p}_{1,i}=-2$, $\eta=0.1$, $\rho=0.3$, and
\[
\Gamma =
\begin{bmatrix}
0 & 0.2 & 0 & 0 & 0.1 & 0 & 0 & 0 \\
0.2 & 0 & 0.1 & 0 & 0 & 0 & 0 & 0 \\
0 & 0.1 & 0 & 0.15 & 0 & 0 & 0 & 0 \\
0 & 0 & 0.15 & 0 & 0.1 & 0 & 0 & 0 \\
0.1 & 0 & 0 & 0.1 & 0 & 0.2 & 0 & 0 \\
0 & 0 & 0 & 0 & 0.2 & 0 & 0.1 & 0 \\
0 & 0 & 0 & 0 & 0 & 0.1 & 0 & 0.2 \\
0 & 0 & 0 & 0 & 0 & 0 & 0.2 & 0
\end{bmatrix}\!\!,\qquad
a=\begin{bmatrix}
 1.0\\
 1.5 \\
 0.8 \\
 1.2 \\
 2.0 \\
 0.9 \\
 1.8 \\
 1.1 
\end{bmatrix}\!\!.
\]
By setting $\rho=10^8$,~\eqref{eq:game-design-relaxed} is solved with 83 LM iterations in 1.057~s starting from the initial guess $p_{1}^{(0)}=\frac{p_{1,\min}+p_{1,\max}}{2}$, $p_{2}^{(0)}=\frac{p_{2,\min}+p_{2,\max}}{2}$, and $x_i^{(0)}=\frac{C}{N}$. The optimal parameters and related GNE are 
\[
    \begin{aligned}
    p_1^\star&=-[1.6174\ 1.8083\ 1.553\  1.6970\  1.9523\ 1.5737\ 1.9009\ 1.6576]^\top\\
    p_2^\star&=0.1990\\
    x^\star(p^\star)&=[0.0882\ 0.1745\ 0.0413\ 0.1325\ 0.2002\ 0.0565\ 0.1945\ 0.1124]^\top
    \end{aligned}
\]
where $\sum_{i=1}^Nx_i^\star(p^\star)\approx 1=C$. The optimal value of the leader's loss function is $J(x^\star(p^\star),p^\star)=0.5637$. We observed that different initializations lead to different GNEs with different values of the leader's loss function, due to the nonconvexity of the optimization problem~\eqref{eq:game-design-relaxed}.

\subsection{LQR game}
We consider a linear system as in~\eqref{eq:linear-system} with $n_x=10$ states and $n_u=10$ inputs, each one controlled by a different agent ($n_i=1$ for all $i=1,\ldots, N$, $N=10$ agents), with  
$A,B\in \mathbb{R}^{10\times 10}$ randomly generated and scaled so that $A$ is unstable with spectral radius equal to 1.1. We consider state-weighting matrices $Q_i=e_ie_i^T$ and $R_i=10$ {for all $i=1,\ldots,N$}.

In~\eqref{eq:approx-LQR-cost}, we set $N_{\textrm LQR}=50$, which is a long-enough horizon used to approximate the infinite-horizon cost, therefore providing a good approximation of the solution of the algebraic Riccati equation associated with the quadruple $((A - B_{-i} K_{-i}), B_i, Q_i, R_i)$ for each agent $i$.

The problem is solved as in~\eqref{eq:NLS} (without parameter $p_0$) in {6.05~s (6} LM iterations),
leading to an asymptotically closed-loop matrix $A - B K^\star$ with spectral radius 0.7525. For comparison, the centralized (cooperative) LQR solution leads to a spectral radius of 0.3965. {Solving the same problem by using the Riccati-based iterations described in~\cite[Section III.B]{NMSM24} leads to the same solution in 0.11~s (7 Riccati iterations).}

\begin{figure}
    \centering
    \includegraphics[width=\hsize]{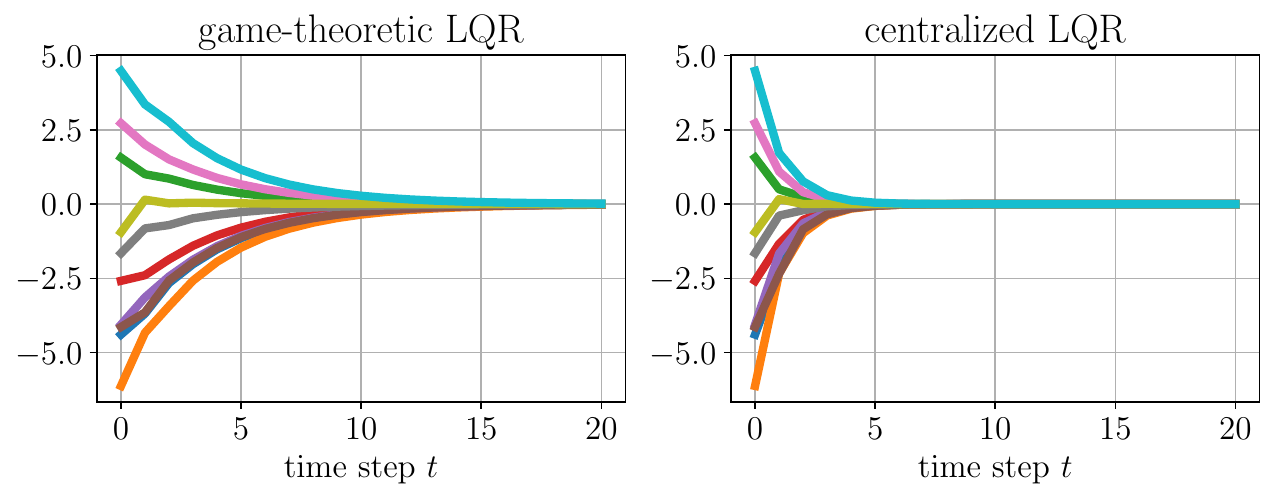}
    \caption{{Game-theoretic vs centralized LQR}}
        \label{fig:example_LQR}
\end{figure}

Figure~\ref{fig:example_LQR} compares the closed-loop step responses of the combined output vector $y(t)=\one^\top x(t)$ from various initial conditions $x(0)$ for the noncooperative LQR gain $K^\star$ and the centralized LQR gain $K_{\rm LQR}$. 

\subsection{Game-theoretic MPC}
\label{sec:mpc-example}
We consider a linear system as in~\eqref{eq:linear-system} with $n_x=6$ states, $n_u=3$ inputs, 
and $n_y=3$ outputs, each one controlled by a different agent ($n_i=1$ for all $i=1,2,3$, $N=3$ agents), with $A,B\in \mathbb{R}^{6\times 3}$ randomly generated and scaled so that $A$ is stable with spectral radius equal to 0.95. The output matrix $C\in\mathbb{R}^{3\times 6}$ is scaled so that the steady-state gain from inputs to outputs is the identity matrix. Each agent $i$ has state-weighting matrix $Q_i=C^\top Q_{y,i} C$, with $Q_{y,i}=e_ie_i^\top$, $Q_{y,i}\in\mathbb{R}^{3\times 3}$, and input-increment weighting $Q_{\Delta u,i}=0.5$ for $i=1,2,3$. The prediction horizon is $T=10$ and the constraint horizon is $T_c=3$. 
{For all $i=1,2,3$}, the input constraints are $0\leq u_i \leq 4$ and the output constraints are $0\leq y_i \leq 5$, with slack penalty $q_{\epsilon,i}=10^3$.

\begin{figure}[t]
    \centering
    \begin{subfigure}[t]{0.48\textwidth}
        \centering
        \includegraphics[width=\linewidth]{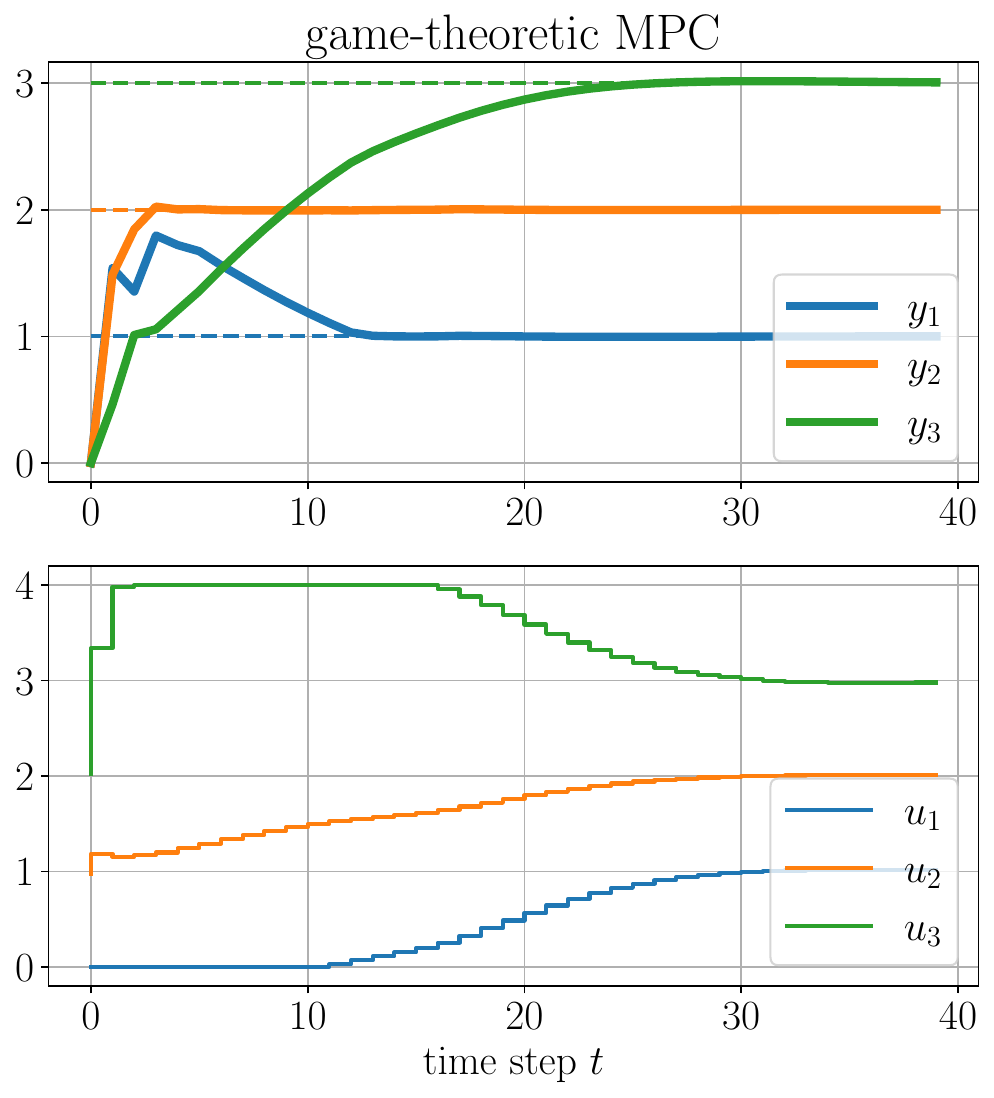}
        \caption{Game-theoretic MPC}
        \label{fig:example_linear_MPC_nash}
    \end{subfigure}\hfill
    \begin{subfigure}[t]{0.48\textwidth}
        \centering
        \includegraphics[width=\linewidth]{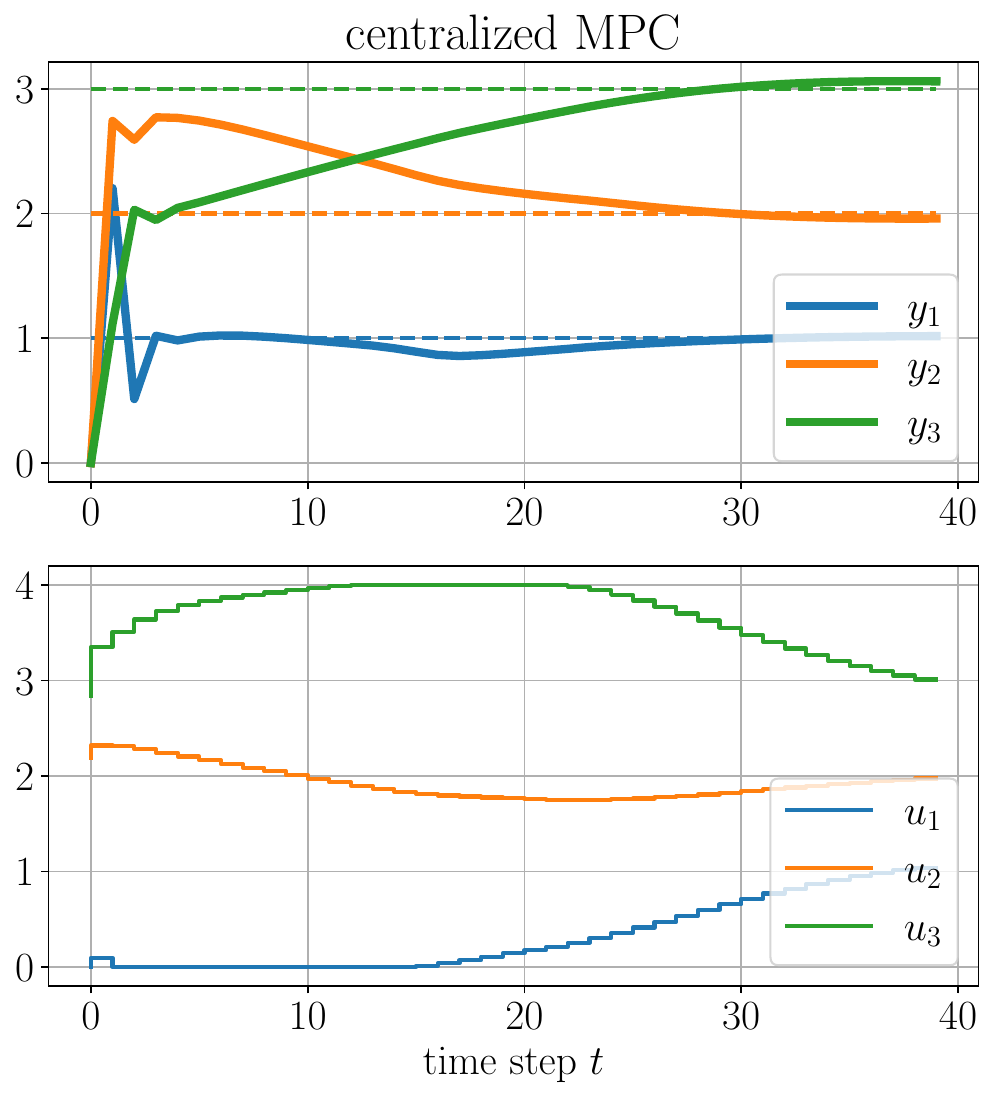}
        \caption{Centralized MPC}
        \label{fig:example_linear_MPC_centralized}
    \end{subfigure}
    \caption{Closed-loop trajectories for linear MPC: comparison between game-theoretic (competitive) and centralized (cooperative) formulations.}
    \label{fig:example_linear_MPC_comparison}
\end{figure}

The noncooperative MPC problem is solved at each time step $t=0,\ldots,T_{\rm sim}-1$, with $T_{\rm sim}=40$, by solving the MILP reformulation~\eqref{eq:GNE-MIP} of the GNEP (without parameter $p$ and objective function~\eqref{eq:game-design-cost}) using the HiGHS solver~\cite{HH18}. The resulting closed-loop trajectories are shown in Figure~\ref{fig:example_linear_MPC_nash}. For comparison, we also show the closed-loop trajectories obtained by solving a cooperative and centralized MPC problem at each time step $t$ by jointly optimizing all agents' moves via the QP solver osQP~\cite{SBGBB20}. 

The CPU time for solving each GNEP ranges between {$4.36$ and $65.62$}~ms (HiGHS MILP), while the CPU time for solving each centralized MPC problem ranges between {$0.57$ and $2.80$}~ms (QP). 

\subsection{{Computing sparse equilibria}}
Consider {a Nash equilibrium problem} with $N=40$ agents, each one controlling a scalar decision variable $x_i$ {and no free parameter (e.g., $p=0$). The} agents are grouped into pairs $(2k-1,2k)$, $k=1,\ldots,20$, with the agents' objectives defined as 
\[
f_i(x) = \bigl(x_{2k-1} - x_{2k}\bigr)^2,\ k=\left\lfloor (i+1)/2 \right\rfloor.
\]
Clearly, if for every agent-pair $k$ the corresponding variables $x_{2k-1} = x_{2k}$, we have a Nash equilibrium $x^\star \in \mathbb{R}^N$. Among the infinitely many Nash equilibria, we want to minimize the game-design objective
\[
J(x,0) = \sum_{i=1}^N (x_i - x_i^{\mathrm{ref}})^2
        + \alpha_1 \|x\|_1
\]
where the components of the reference vector $x^{\mathrm{ref}}_i = \frac{\lceil(i+1)/2\rceil}{10}$, $i=1,\ldots,N$,
and $\alpha_1 > 0$ promotes sparsity in the equilibrium solution. 

\begin{figure}[t]
    \centering
    \includegraphics[width=0.7\textwidth]{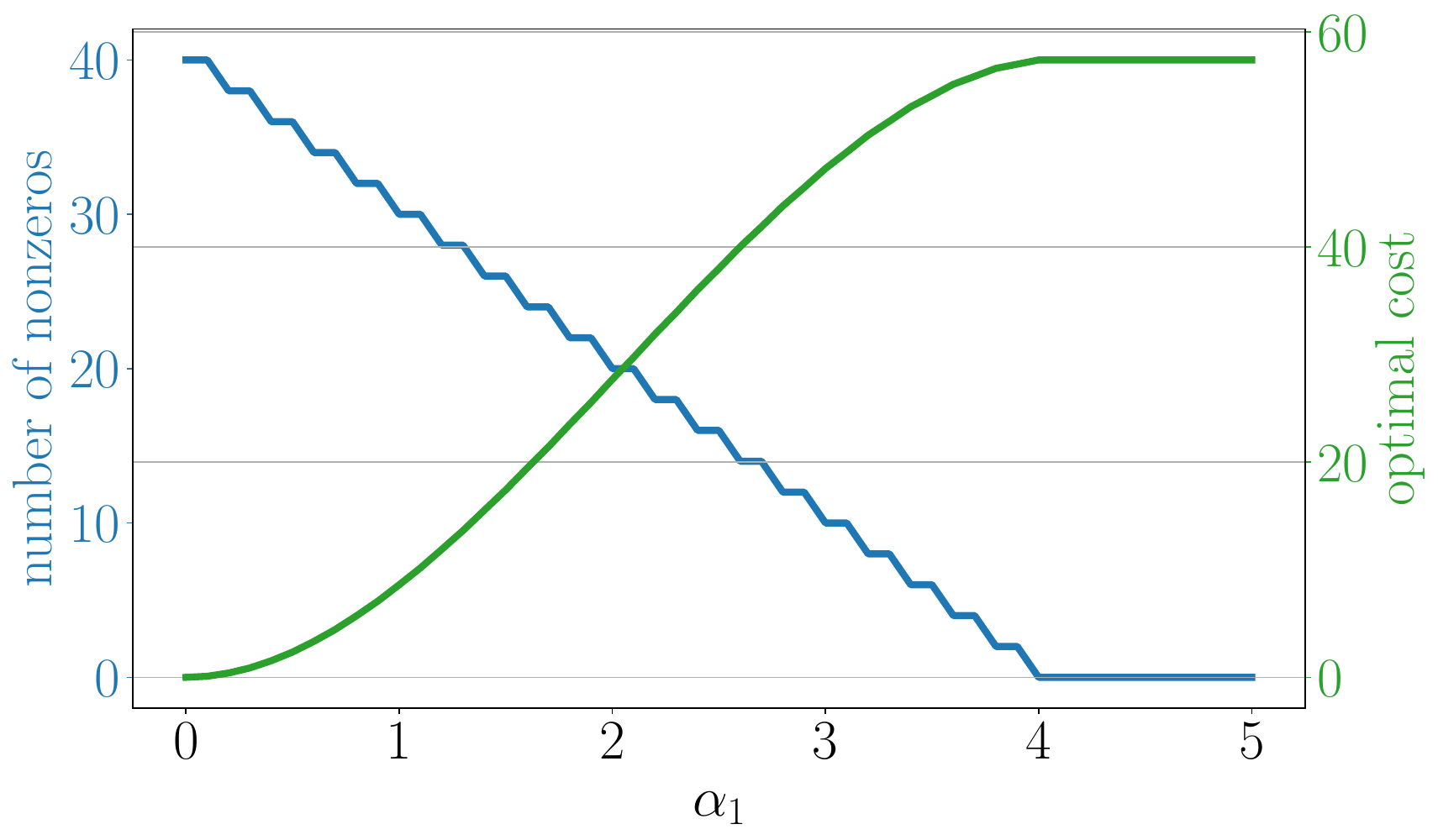} 
    \caption{Number of nonzeros in $x^\star$ and optimal cost $J(x^\star,0)$ as a function of the $\ell_1$-regularization parameter $\alpha_1$.}
    \label{fig:sparse_GNE}
\end{figure}

Figure~\ref{fig:sparse_GNE} shows the number of nonzeros in the computed {Nash equilibrium} $x^\star$ and the optimal cost $J(x^\star,0)$ as a function of the $\ell_1$-regularization parameter $\alpha_1$, obtained by solving Problem~\eqref{eq:GNE-relaxed-L1} with $\rho=10^4$ {and $P=\{0\}$}. The CPU time is 1.4{350} s on the first run due to JAX compilation time, and {29.9 ms} on average on the subsequent runs. As expected, as $\alpha_1$ increases, the number of zeros in $x^\star$ increases, leading to sparser Nash equilibria, at the expense of a larger optimal cost $J(x^\star,0)$. Clearly, due to the equilibrium condition $x_{2k-1}^\star = x_{2k}^\star$, the number of nonzeros in $x^\star$ can only decrease by two at a time as $\alpha_1$ increases.

\begin{figure}[t]
    \centering
    \includegraphics[width=0.65\textwidth]{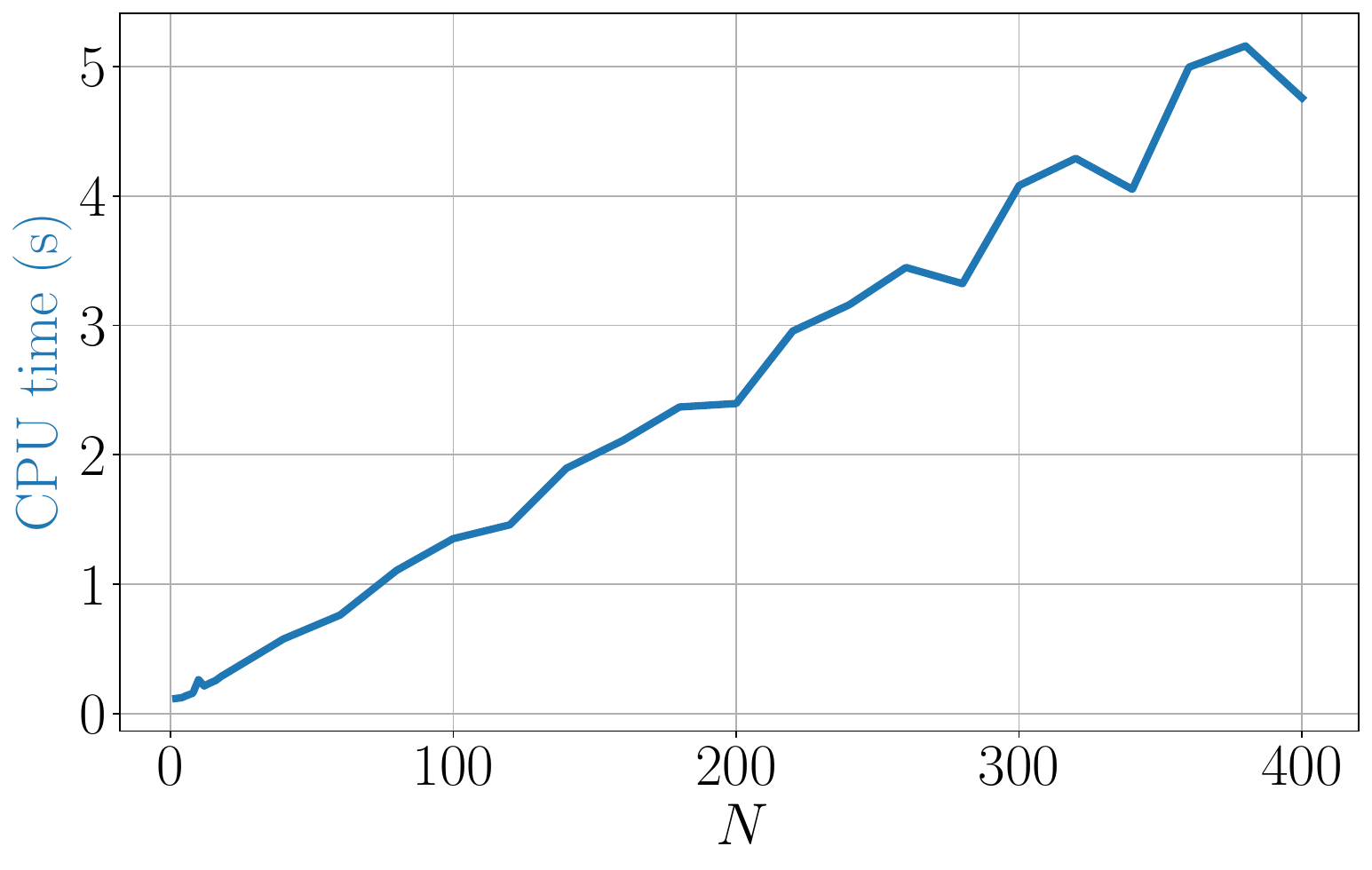} 
    \caption{CPU time to compute the Nash equilibrium as a function of the number $N$ of agents.}
    \label{fig:sparse_GNE_time}
\end{figure}

Figure~\ref{fig:sparse_GNE_time} shows the CPU time required to compute the sparse solution as a function of the number $N$ of agents for $\alpha_1=2$ using L-BFGS-B. The CPU time grows roughly linearly with $N$.

\section*{Acknowledgments}
We thank the anonymous reviewers for their constructive comments and suggestions.

\section{Conclusions}
In this paper, we have proposed different simple methods for designing and computing generalized Nash equilibria in noncooperative games with local and shared constraints. We have presented a mixed-integer linear or quadratic programming reformulation for computing all generalized Nash equilibria of a linear-quadratic game, as well as a nonlinear least-squares approach for computing generalized Nash equilibria of a rather general class of nonlinear games. We have illustrated the proposed methods on several numerical examples, including linear-quadratic games, game-theoretic LQR and MPC control, inverse games, and sparsity-promoting GNE computations. We believe that the proposed methods and the associated open-source library can be a useful tool for researchers and practitioners working in the field of game-theory and game-theoretic control and its applications to engineering and economic systems.


\begin{thebibliography}{10}

\bibitem{AS20}
D.~Aussel and A.~Svensson.
\newblock A short state of the art on multi-leader-follower games.
\newblock In S.~Dempe and J.~Zemkoho, editors, {\em Bilevel Optimization}, volume 161, pages 53--76. Springer, 2020.

\bibitem{Bau16}
D.~Bauso.
\newblock {\em Game Theory with Engineering Applications}.
\newblock SIAM, 2016.

\bibitem{BYGP22}
G.~Belgioioso, P.~Yi, S.~Grammatico, and L.~Pavel.
\newblock Distributed generalized {Nash} equilibrium seeking: An operator-theoretic perspective.
\newblock {\em IEEE Control Systems Magazine}, 42(4):87--102, August 2022.

\bibitem{Bem04b}
A.~Bemporad.
\newblock {\em Hybrid Toolbox -- {U}ser's Guide}.
\newblock January 2004.
\newblock \url{http://cse.lab.imtlucca.it/~bemporad/hybrid/toolbox}.

\bibitem{Bem25}
A.~Bemporad.
\newblock An {L-BFGS-B} approach for linear and nonlinear system identification under $\ell_1$ and group-lasso regularization.
\newblock {\em IEEE Transactions on Automatic Control}, 70(7):4857--4864, 2025.
\newblock code available at \url{https://github.com/bemporad/jax-sysid}.

\bibitem{BMR04}
A.~Bemporad, M.~Morari, and N.L. Ricker.
\newblock {\em Model Predictive Control Toolbox for {MATLAB} -- User's Guide}.
\newblock The Mathworks, Inc., 2004.
\newblock \url{http://www.mathworks.com/access/helpdesk/help/toolbox/mpc/}.

\bibitem{BK21}
E.~B\"{o}rgens and C.~Kanzow.
\newblock {ADMM}-type methods for generalized {Nash} equilibrium problems in {Hilbert} spaces.
\newblock {\em SIAM Journal on Optimization}, 31(1):377--403, 2021.

\bibitem{JAX}
J.~Bradbury, R.~Frostig, P.~Hawkins, M.J. Johnson, C.~Leary, D.~Maclaurin, G.~Necula, A.~Paszke, J.~Vander{P}las, S.~Wanderman-{M}ilne, and Q.~Zhang.
\newblock {JAX}: composable transformations of {P}ython+{N}um{P}y programs, 2018.

\bibitem{BFH19}
L.F. Bueno, G.~Haeser, and F.N. Rojas.
\newblock Optimality conditions and constraint qualifications for generalized {Nash} equilibrium problems and their practical implications.
\newblock {\em SIAM Journal on Optimization}, 29(1):31--54, 2019.

\bibitem{BLNZ95}
R.H. Byrd, P.~Lu, J.~Nocedal, and C.~Zhu.
\newblock A limited memory algorithm for bound constrained optimization.
\newblock {\em SIAM Journal on Scientific Computing}, 16(5):1190--1208, 1995.

\bibitem{LJWA20}
H.~Le Cadre, P.~Jacquot abd C.~Wan, and C.~Alasseur.
\newblock Peer-to-peer electricity market analysis: From variational to generalized {Nash} equilibrium.
\newblock {\em European Journal of Operational Research}, 282:753--771, 2020.

\bibitem{CL96}
T.F. Coleman and Y.~Li.
\newblock An interior trust region approach for nonlinear minimization subject to bounds.
\newblock {\em SIAM Journal on Optimization}, 6(2):418--445, 1996.

\bibitem{DFKS11}
A.~Dreves, F.~Facchinei, C.~Kanzow, and S.~Sagratella.
\newblock On the solution of the {KKT} conditions of generalized {Nash} equilibrium problems.
\newblock {\em SIAM Journal on Optimization}, 21(3):1082--1108, 2011.

\bibitem{DS16}
A.~Dreves and N.~Sudermann-Merx.
\newblock Solving linear generalized {Nash} equilibrium problems numerically.
\newblock {\em Optimization Methods and Software}, 31(5):1036--1063, 2016.

\bibitem{FB25}
F.~Fabiani and A.~Bemporad.
\newblock An active learning method for solving competitive multi-agent decision-making and control problems.
\newblock {\em IEEE Transactions on Automatic Control}, 70(4):2374--2389, 2025.
\newblock code availble at \url{https://github.com/bemporad/gnep-learn}.

\bibitem{FFK98}
F.~Facchinei, A.~Fischer, and C.~Kanzow.
\newblock Regularity properties of a semismooth reformulation of variational inequalities.
\newblock {\em SIAM Journal on Optimization}, 8(3):850--869, 1998.

\bibitem{FFP07}
F.~Facchinei, A.~Fischer, and V.~Piccialli.
\newblock Generalized {Nash} equilibrium problems and {Newton} methods.
\newblock {\em Mathematical Programming}, 117(1-2):163--194, July 2007.

\bibitem{FZ10}
F.~Facchinei and C.~Kanzow.
\newblock Generalized {Nash} equilibrium problems.
\newblock {\em Annals of Operations Research}, 175:177--211, 2010.

\bibitem{Fischer1992}
A.~Fischer.
\newblock A special {Newton}-type optimization method.
\newblock {\em Optimization}, 24(3--4):269--284, 1992.

\bibitem{BFSS25}
B.~Franci, F.~Fabiani, M.~Schmidt, and M.~Staudigl.
\newblock A {Gauss–Seidel} method for solving multi-leader-multi-follower games.
\newblock In {\em European Control Conference}, pages 2775--2780, 2025.

\bibitem{GKW25}
G.~Graser, T.~Kreimeier, and A.~Walther.
\newblock Solving linear generalized {Nash} games using an active signature method.
\newblock {\em Optimization Methods and Software}, pages 1--24, 2025.

\bibitem{GLY15}
L.~Guo, G.-H. Lin, and J.J. Ye.
\newblock Solving mathematical programs with equilibrium constraints.
\newblock {\em Journal of Optimization Theory and Applications}, 166:234--256, 2015.

\bibitem{Gurobi}
{Gurobi Optimization, LLC}.
\newblock {Gurobi Optimizer Reference Manual}, 2024.

\bibitem{HBLD22}
S.~Hall, G.~Belgioioso, D.~{Liao-McPherson}, and F.~D\"orfler.
\newblock Receding horizon games with coupling constraints for demand-side management.
\newblock In {\em Proc. IEEE 61st Conference on Decision and Control}, page 3795–3800, December 2022.

\bibitem{HB26}
S.~Hall and M.~Bemporad.
\newblock Solving multiparametric generalized {Nash} equilibrium problems and explicit game-theoretic model predictive control.
\newblock 2025.
\newblock available on arXiv at \url{https://arxiv.org/abs/2512.05505}. Code available at \url{https://github.com/bemporad/nash_mpqp}.

\bibitem{HDB01a}
W.P.M.H. Heemels, B.~De Schutter, and A.~Bemporad.
\newblock Equivalence of hybrid dynamical models.
\newblock {\em Automatica}, 37(7):1085--1091, July 2001.

\bibitem{HH18}
Q.~Huangfu and J.A.J. Hall.
\newblock Parallelizing the dual revised simplex method.
\newblock {\em Mathematical Programming Computation}, 10(1):119--142, 2018.

\bibitem{KS16}
C.~Kanzow and D.~Steck.
\newblock Augmented {Lagrangian} methods for the solution of generalized {Nash} equilibrium problems.
\newblock {\em SIAM Journal on Optimization}, 26(4):2034--2058, 2016.

\bibitem{LSM22}
S.~{Le Cleac'h}, M.~Schwager, and Z.~Manchester.
\newblock {ALGAMES}: a fast augmented {Lagrangian} solver for constrained dynamic games.
\newblock {\em Autonomous Robots}, 46(1):201--215, 2022.

\bibitem{Lev44}
K.~Levenberg.
\newblock A method for the solution of certain non-linear problems in least squares.
\newblock {\em Quarterly of Applied Mathematics}, 2(2):164--168, 1944.

\bibitem{LK24}
M.~Liu and I.V. Kolmanovsky.
\newblock Input-to-state stability of {Newton} methods in {Nash} equilibrium problems with applications to game-theoretic model predictive control.
\newblock {\em arXiv preprint arXiv:2412.06186}, 2024.

\bibitem{LPR96}
Z.-Q. Luo, J.-S. Pang, and D.~Ralph.
\newblock {\em Mathematical programs with equilibrium constraints}.
\newblock Cambridge University Press, 1996.

\bibitem{Mar63}
D.W. Marquardt.
\newblock An algorithm for least-squares estimation of nonlinear parameters.
\newblock {\em Journal of the Society for Industrial and Applied Mathematics}, 11(2):431--441, 1963.

\bibitem{NW06}
J.~Nocedal and S.J. Wright.
\newblock {\em Numerical Optimization}.
\newblock Springer, 2 edition, 2006.

\bibitem{NMSM24}
B.~Nortmann, A.~Monti, M.~Sassano, and T.~Mylvaganam.
\newblock {Nash} equilibria for linear quadratic discrete-time dynamic games via iterative and data-driven algorithms.
\newblock {\em IEEE Transactions on Automatic Control}, 69(10):6561--6575, 2024.

\bibitem{PF05}
J.-P. Pang and M.~Fukushima.
\newblock Quasi-variational inequalities, generalized {Nash} equilibria, and multi-leader-follower games.
\newblock {\em Computational Management Science}, 2(1):21--56, January 2005.

\bibitem{SBGBB20}
B.~Stellato, G.~Banjac, P.~Goulart, A.~Bemporad, and S.~Boyd.
\newblock {OSQP}: An operator splitting solver for quadratic programs.
\newblock {\em Mathematical Programming Computation}, 12:637--672, 2020.

\bibitem{TK18}
T.~Tatarenko and M.~Kamgarpour.
\newblock Learning generalized {Nash} equilibria in a class of convex games.
\newblock {\em IEEE Transactions on Automatic Control}, 64(4):1426--1439, 2018.

\bibitem{TSN21}
T.~Tatarenko, W.~Shi, and A.~Nedi\'{c}.
\newblock Geometric convergence of gradient play algorithms for distributed {Nash} equilibrium seeking.
\newblock {\em IEEE Transactions on Automatic Control}, 66(11):5342--5353, 2021.

\bibitem{WLMCMWW21}
Z.~Wang, F.~Liu, Z.~Ma, Y.~Chen, W.~Wei, and Q.~Wu.
\newblock Distributed generalized {Nash} equilibrium seeking for energy sharing games in prosumers.
\newblock {\em IEEE Transactions on Power Systems}, 36(6):3973--3986, September 2021.

\bibitem{YP19}
P.~Yi and L.~Pavel.
\newblock An operator splitting approach for distributed generalized {Nash} equilibria computation.
\newblock {\em Automatica}, 102:111--121, 2019.

\end{thebibliography}
\end{document}